\documentclass{aa}
\pdfoutput=1
\usepackage[varg]{txfonts}
\usepackage{graphicx}
\usepackage{natbib}
\bibpunct{(}{)}{;}{a}{}{,}

\begin{document}
\title{On the frequency of star-forming galaxies in the vicinity of powerful AGNs: The case of SMM J04135+10277}
\author{J. Fogasy \inst{1}
	\and K. K. Knudsen \inst{1}
	\and C. D. P. Lagos \inst{2, 3}
	\and G. Drouart \inst{1, 4} 
        \and V. Gonzalez-Perez \inst{5, 6}}
\institute{Department of Earth and Space Sciences, Chalmers University of Technology, Onsala Space Observatory, 439 92 Onsala, Sweden\\ \email{judit.fogasy@chalmers.se}
	\and International Centre for Radio Astronomy Research, 7 Fairway, Crawley, 6009, Perth, WA, Australia 
        \and Australian Research Council Centre of Excellence for All-sky Astrophysics (CAASTRO), 44 Rosehill Street Redfern, NSW 2016, Australia 
        \and Curtin University, Department of Physics and Astronomy, Kent Street, Bentley, 6102, Perth, WA, Australia
        \and Institute for Computational Cosmology, Department of Physics, University of Durham, South Road, Durham DH1 3LE, UK
        \and Institute of Cosmology \& Gravitation, University of Portsmouth, Dennis Sciama Building, Portsmouth PO1 3FX, UK}
\date{Received , 2016; accepted , 2016}
\abstract 
{In the last decade several massive molecular gas reservoirs were found $<100$\,kpc distance
from active galactic nuclei (AGNs), residing in gas-rich companion galaxies. 
The study of AGN--gas-rich companion systems opens the opportunity to determine whether the stellar mass of massive local galaxies was formed in their host after a merger event or outside of their host galaxy in a close starbursting companion and later incorporated via mergers.}
{Our aim is to study the quasar--companion galaxy system of SMM\,J04135+10277 ($z=2.84$) and investigate the expected frequency of quasar--starburst galaxy pairs at high redshift using a cosmological galaxy formation model.}
{We use archive data and new APEX ArTeMiS data to construct and model the spectral energy distribution of SMM\,J04135+10277 in order to determine its properties. We also carry out a comprehensive analysis of the cosmological galaxy formation model {\sc galform} with the aim of characterising how typical the system of SMM\,J04135+10277 is and whether quasar--star-forming galaxy pairs may constitute an important stage in galaxy evolution. Finally, we compare our results to observations found in the literature at both large and small scales (1 Mpc--100 kpc).}
{The companion galaxy of SMM\,J04135+10277 is a heavily dust-obscured starburst galaxy with a median star formation rate (SFR) of $700\,\rm{M_{\odot}\,yr^{-1}}$, median dust mass of $5.1\times 10^9\,\rm{M_{\odot}}$ and median dust luminosity of $\textrm 9.3 \times 10^{12}\, \rm{L_{\odot}}$.
Our simulations, performed at $z=2.8$, suggest that SMM\,J04135+10277 is not unique. In fact, at a distance of $<100$~kpc, 22\% of 
our simulated quasar sample have at least one companion galaxy of a stellar mass $>10^8\,\rm{M_{\odot}}$, and 0.3\% have at least one highly star-forming companion ($\rm{SFR}>100\ \rm{M_{\odot}\,yr^{-1}}$).}
{Our results suggest that quasar--gas-rich companion galaxy systems are common phenomena in the early Universe and the high incidence of companions makes the study of such systems crucial to understand the growth and hierarchical build-up of galaxies and black holes.}

\maketitle

\section{Introduction}
One of the many fascinating questions of astronomy is how galaxies evolve and grow and in particular how local, massive galaxies formed. According to the correlation found between the central super-massive black hole (SMBH) mass and bulge mass or velocity dispersion of the bulge, the SMBH and its host galaxy evolve together \citep[e.g.][]{1998AJ....115.2285M, 2000ApJ...539L..13G, 2001ApJ...547..140M, 2002ApJ...574..740T, 2003ApJ...589L..21M, 2004ApJ...604L..89H, 2013ApJ...764..184M}. The exact origin of the $M_{\rm BH}-M_{\rm bulge}$ or $M_{\rm BH}-\sigma$ is not completely understood but it has its root in the fuelling and termination of star formation and black hole accretion.

The stellar mass of galaxies can grow either by quiescent star formation through gas accretion  onto the galaxies or by starburst mode, where the star formation is likely triggered by interactions and mergers of gas-rich systems \citep[e.g.][]{2010ApJ...714L.118D, 2010MNRAS.407.2091G, 2011ApJ...739L..40R}. Considering the merger scenario, it is possible that in a fraction of local, massive galaxies part of their stellar mass was formed outside of their host galaxy in a close starburst companion galaxy and was later incorporated via mergers.

The study of high-redshift quasars and radio galaxies is essential to understand the formation and evolution of galaxies, since their SMBH undergoes a rapid growth \citep[e.g.][and references therein]{2012NewAR..56...93A}. Moreover, one-third of optically selected quasars show signs of intense star formation through the presence of cold dust \citep{2001ApJ...555..625C, 2001A&A...374..371O} and several high-redshift radio galaxies have luminous CO emission regions \citep[e.g.][]{2000ApJ...528..626P, 2005A&A...430L...1D, 2005ApJ...621L...1K, 2016A&A...586A.124G}. In some of these sources high-resolution observations revealed that the molecular gas reservoirs were associated with close companion galaxies and not with the host galaxies of the active galactic nuclei (AGNs). For instance, the quasar BR 1202-0725 ($z=4.7$) has a gas-rich companion galaxy at 25 kpc distance \citep{2002AJ....123.1838C, 2012A&A...545A..57S, 2013A&A...559A..29C}, while the high-$z$ radio galaxy (HzRG) B3 J2330+3927 ($z=3.09$) has potentially two gas-rich companions at a distance of 30 kpc \citep{2012MNRAS.425.1320I}.

Another interesting source, and the subject of this paper, is SMM J04135+10277 ($z=2.837$), which is a submillimetre bright type 1 quasar discovered  by \citet{2003A&A...411..343K} with the Submillimetre Common-User Bolometer Array (SCUBA) mounted on the James Clerk Maxwell Telescope (JCMT). The quasar is gravitationally magnified by the foreground galaxy cluster A478 with a gravitational magnification factor of 1.3 \citep{2003A&A...411..343K}. CO(3-2) and CO(1-0) observations showed that this galaxy has one of the most massive molecular gas reservoirs seen in the high-$z$ Universe ($M_{\rm H_2} \sim 10^{11}$\,M$_\odot$, $L'_{\rm CO(1-0)}=1.84\times 10^{11}\, \rm{K\ km\ s^{-1}\ pc^2}$; \citealt{2004ApJ...609...61H, 2011ApJ...739L..32R}). Moreover, the \textit{Spitzer Space Telescope} detected significant PAH emission, corresponding to a star formation rate (SFR) of 1680 $\rm {M_{\odot}\,yr^{-1}}$ \citep{2008ApJ...684..853L}.

However, recent high-resolution observations using the Combined Array for Research in Millimeter-wave Astronomy (CARMA) interferometer revealed that the CO(3-2) emission is not centred on the quasar itself but on an optically undetected galaxy $5.2\arcsec$ (41.5 kpc) away from the quasar position \citep{2013ApJ...765L..31R}. The molecular gas emission region is distributed over $1.7\arcsec$ (13 kpc) along its major axis but remains unresolved  down to $1.2\arcsec$ on its minor axis. While the submm continuum emission is unresolved in the $15\arcsec$ beam of SCUBA, based on the CO(3-2) detection it is most likely coming from the gas-rich companion galaxy as well.
It has been suggested that this system could represent an early stage of a wet-dry merger event, given the separation between the quasar and its companion galaxy and the molecular gas distribution \citep{2013ApJ...765L..31R}.

In this paper, we study the companion galaxy of the quasar SMM J04135+10277 by constructing and modelling its spectral energy distribution (SED). In addition to the observational data, we carried out galaxy formation simulations to investigate the expected frequency of such quasar--gas-rich galaxy pairs.
In Section 2 we present our observations and the archive data we used. In Section 3 we present our observational results and the SED modelling of the companion galaxy. In Section 4 we describe the details and results of the galaxy formation simulations. In Section 5 we discuss the observational and simulational findings and compare them to other studies found in the literature. In Section 6 we present our conclusions.

Throughout this paper we adopt WMAP7 cosmology with $H_{0}=70.4\ \rm{km \ s^{-1} Mpc^{-1}}$, $\rm \Omega_{m}=0.272$ and $\rm \Omega_{\Lambda}=0.728$ \citep{2011ApJS..192...18K}.
\section{Observation and data reduction}
\subsection{APEX ArTeMiS data}
The Atacama Pathfinder EXperiment (APEX) ArTeMiS\footnote{Architectures de bolometres pour des Telescopes a grand champ de vue dans le domaine sub-Millimetrique au Sol} observation of SMM J04135+10277 was obtained during four nights between 25 September and 5 October 2014.
ArTeMiS is an ESO PI bolometer camera operating in the submm range (200, 350, 450 $\mu$m) on APEX. The camera consists of Si:P:B bolometeres arranged in 16$\times$18 sub-arrays \citep{2014SPIE.9153E..05R}.
The total observing time at 350 $\mu \rm m$ was 14.3 hours in spiral raster mapping mode. The source was observed under good weather conditions with precipitable water vapour between 0.3-0.5 mm. The resulting images have an angular resolution of $8\arcsec$. The data were reduced using the ArTeMiS data reduction package for IDL provided by the ArTeMis team  following the same method as \citet{2008A&A...490L..27A}. The data reduction includes: reduction of skydip observations to obtain opacity information for each scan; extinction correction, median baseline subtraction, correlated median noise subtraction, unit conversion (Jy/beam), and map building for each scan; and a combination of reduced scans. HLTau was observed as flux calibrator, and the uncertainty of flux calibration is $\sim 25 \%$.
\subsection{Optical and infrared archive data}
\subsubsection{Optical data}
Part of the optical data was obtained from the Hubble Legacy Archive\footnote{\url{http://hla.stsci.edu}} (HLA), including pipeline processed, calibrated, combined Wide-Field Planetary Camera 2 (WFPC2; \citealt{1994ApJ...435L...3T}) F555W, F606W and F814W  images (project ID:7337, PI: A. C. Fabian; Project ID:8301, PI: A. Edge). These HLA processed images are astrometrically corrected by matching point sources to the 2 Micron All Sky Survey (2MASS) catalogue. The typical astrometric accuracy of the HLA-produced images is $\sim0.3\arcsec$ in each coordinate. The absolute photometry is good within 0.03 mag according to the HLA website.

Additionally, we used the Canada France Hawaii Telescope\footnote{\url{http://www.cadc-ccda.hia-iha.nrc-cnrc.gc.ca/en/}} (CFHT) MegaPipe processed MegaPrime/MegaCam \textit{g}- and \textit{r}-band data. The MegaPipe processed images have astrometric accuracy of $0.15\arcsec$ relative to external reference frames and $0.04\arcsec$ internally. The photometric calibration of MegaPipe images is good within 0.03 mag \citep{2008PASP..120..212G}. We combined the available \textit{g}- and \textit{r}-band data in \texttt{IRAF} using the \texttt{imcombine} task.

The companion galaxy is not detected in any of the optical images, only the quasar, thus we measured the noise around the position of the companion galaxy and calculated 5$\sigma$ upper limits. We applied Galactic extinction correction to the optical data ($A_{V}=1.44$ mag; \citealt{2011ApJ...737..103S}). The resulting flux densities are given in Table \ref{table1}.

\subsubsection{Infrared data}
The mid-infared part of the SED is covered by \textit{Spitzer} Infrared Array Camera (IRAC; \citealt{2004ApJS..154...10F}) Basic Calibrated Data (BCD) and post-BCD at 3.6-8.0 $\mu$m (project ID: 3720, PI: J. Huang) and Multiband Imaging Photometer for Spitzer (MIPS; \citealt{2004ApJS..154...25R}) BCD at 24 $\mu$m (project ID: SMMJ0413/20631, PI: K. Knudsen). The far-infrared (FIR) data were obtained from JCMT SCUBA 450 $\mu$m and 850 $\mu$m observations \citep{2003A&A...411..343K}.

The pointing accuracy of the IRAC images is better than $0.5\arcsec$, while the MIPS 24 $\mu$m channel has an absolute pointing accuracy of $1.4\arcsec$. The uncertainty of the flux calibration is better than $10\%$ for all the IRAC channels and for the MIPS 24 $\mu$m band.
In case of the 5.8 $\mu$m and 8.0 $\mu$m IRAC channels and the MIPS 24 $\mu$m data we used the MOsaicker and Point source EXtractor package (MOPEX; \citealt{2005PASP..117.1113M})\footnote{Publicly available at \url{http://irsa.ipac.caltech.edu/data/SPITZER/docs/dataanalysistools/tools/mopex/}} to reduce and combine the BCD images.

We calculated the fluxes by aperture photometry with an aperture of $2.4\arcsec$ for each IRAC channel and applied aperture correction according to the IRAC Instrument Handbook\footnote{\url{http://irsa.ipac.caltech.edu/data/SPITZER/docs/irac/iracinstrumenthandbook/}}. The Astronomical Point source EXtractor (APEX) module of MOPEX was used to fit the point response function of the source and measure the flux density for the MIPS 24 $\mu$m band.

The resulting flux densities are listed in Table \ref{table1}. The flux uncertainties include the statistical errors and the flux calibration errors, added together in quadrature.

\section{Results and analysis}
\subsection{Photometry results}
\begin{table}
\caption{Photometry results of SMM J04135+10277\tablefootmark{a}          
\label{table1}      }
\centering          
\begin{tabular}{l c  }
\hline\hline       
\rule{0pt}{2.5ex}Observing band & $S_{\textrm{Companion}}\tablefootmark{b}\ [\mu \textrm{Jy}]$\\ 
\hline                    
\rule{0pt}{2.5ex}MegaPrime G & <0.05 \\
WFPC2 F555W &  <0.06 \\
WFPC2 F606W & <0.07  \\
MegaPrime R  & <0.06  \\
WFPC2 F814W &  <0.06  \\
IRAC 3.5 $\mu$m  &31.7$\pm$3.2 \\
IRAC 4.5 $\mu$m & 44.1$\pm$4.4 \\
IRAC 5.8 $\mu$m & 54.2$\pm$5.7 \\
IRAC 8.0 $\mu$m  & 80.1$\pm$8.4 \\
MIPS 24 $\mu$m & <356\\
ArTeMiS 350 $\mu$m & <100 mJy\\
SCUBA 450 $\mu$m\tablefootmark{c} & 55 mJy$\pm$17 mJy\\
SCUBA 850 $\mu$m\tablefootmark{d} & 25 mJy$\pm$2.8 mJy \\
\hline                  
\end{tabular}
\tablefoot{
\tablefoottext{a}{{The fluxes are not corrected for gravitational lensing magnification.}}
\tablefoottext{b}{In case of non-detection 5$\sigma$ upper limit is shown.}
\tablefoottext{c,d}{From \citet{2003A&A...411..343K}.}}
\end{table}

\begin{figure}
\centering
\includegraphics[width=0.89\hsize]{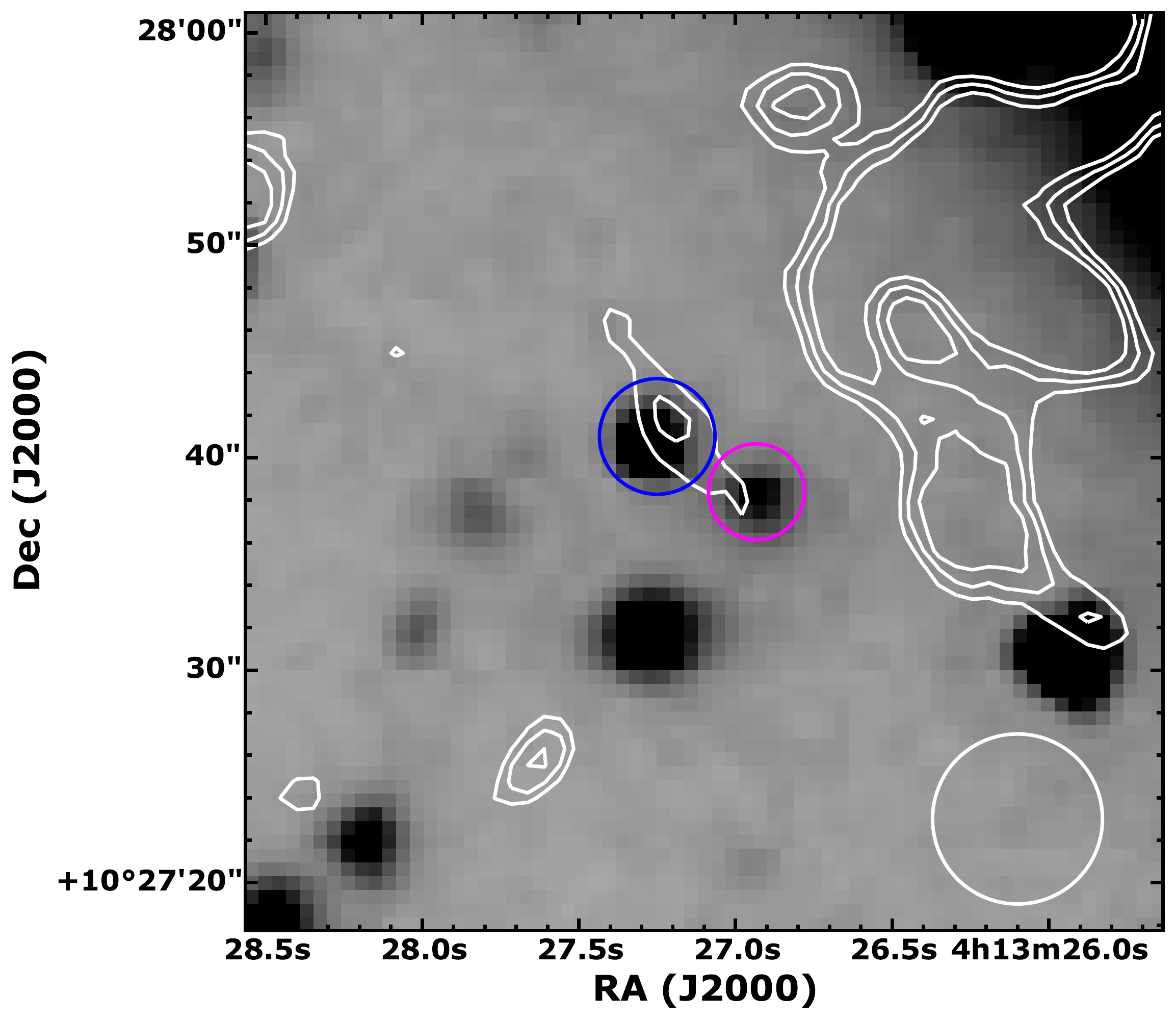}
\caption{ArTeMiS 350 $\mu$m contour map of SMM J04135+10277 overlayed on the \textit{Spitzer} IRAC $4.5\,\mu$m image in the case of Gaussian smoothing with a kernel radius of 5 pixels. The blue and magenta circles denote the quasar and its companion galaxy, respectively. The contour levels are at 30, 35, and 40 mJy. The $8\arcsec$ beam size is indicated in the bottom right corner.}
\label{fig1}
\end{figure}

The final ArTeMiS 350 $\mu$m map has an rms of 20 mJy after Gaussian smoothing (Fig.~\ref{fig1}). 
Even though there is a marginal detection at the position of SMM J04135+10277, we treat it as a non-detection given the increased rms around the source and we calculate a 5$\sigma$ upper limit.

The companion galaxy is only detected in the IRAC bands, and for the optical bands we calculated 5$\sigma$ upper limits. The non-detection of the companion galaxy in the optical bands indicates high visual extinction presented in the galaxy ($A_{V}$).

As the resolution of MIPS at 24 $\mu$m is $6\arcsec$ and the emission is blended over the position of the quasar and the companion galaxy, both of these sources might have contribution to the total 24 $\mu$m emission. Based on the flux ratio of the quasar and the companion galaxy in the IRAC channel 3 and 4, which is 0.38 and 0.23, respectively, the estimated flux density of the companion galaxy is up to 25\% of the total 24 $\mu$m emission, which gives an upper limit of 356 $\mu$Jy.

The CARMA detected huge molecular gas reservoir associated with the companion galaxy  and  the empirical relationship between $L_{\rm FIR}$--$L'_{\rm CO(1-0)}$ for high-$z$, star-forming galaxies \citep{2013ARA&A..51..105C} indicates that the companion galaxy must be very bright at FIR wavelengths ($L_{\rm FIR}\sim 10^{13}\,\rm{L_{\odot}}$). Therefore in case of the SCUBA 450 $\mu$m and 850 $\mu$m bands we make the assumption that all the continuum emission is coming from the companion galaxy, rather than the host galaxy of the quasar.

\subsection{Spectral energy distribution modelling}
To model the SED of the companion galaxy we used the \textsc{magphys} code with high-$z$ extensions \citep{2008MNRAS.388.1595D,2015ApJ...806..110D}\footnote{Publicly available at \url{http://www.iap.fr/magphys.}}.
The \textsc{magphys} code has the advantage of simultaneously fitting the ultraviolet-to-mm SED of galaxies. The \textsc{magphys} code models the star formation history, which is parameterised as a continuous delayed exponential function; the dust attenuation, using the two-component model of \citet{2000ApJ...539..718C}; and the metallicity of galaxies, and this code combines the stellar emission with the dust emission using an energy balance technique. A Bayesian method is used to compare the SED models with the observed SED of the galaxies. Although the fits are performed excluding the optical upper limits, these are consistent with the best fits we obtain.

\begin{figure}
   \centering
\resizebox{\hsize}{!}{\includegraphics[angle=90]{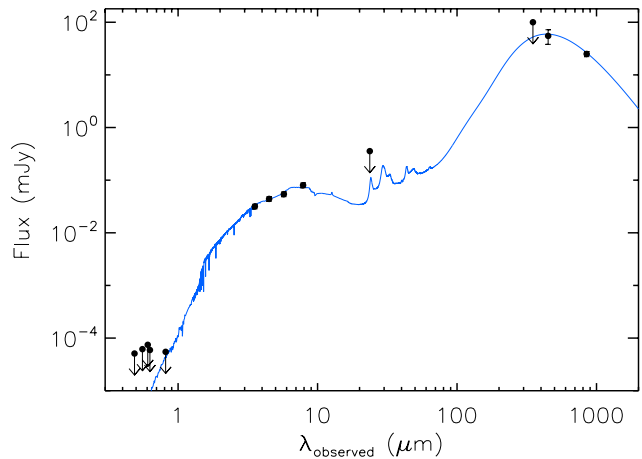}}
     \caption{Spectral energy distribution of the companion galaxy. The blue curve indicates the dust attenuated \textsc{magphys} model of the SED.}
       \label{fig2}
   \end{figure}

In Table \ref{table2} we summarise the median and $16^{\textrm{th}}$ and $84^{\textrm{th}}$ percentile ranges of the fitted parameters. Based on the \textsc{magphys} fit the median visual extinction is high in the companion galaxy ($A_{V}=2.8$ mag). The predicted median stellar mass and SFR of the companion galaxy is  $ M_{*}=7.4 \times 10^{11}\ \rm{M_{\odot}}$ and $ 700\ \rm{M_{\odot}\,yr^{-1}}$, respectively.
The median dust mass of the galaxy is $ 5.1 \times 10^{9}\,\rm {M_{\odot}}$ with an averaged, luminosity-weighted dust temperature of 34 K and dust luminosity of $ 9.3 \times 10^{12}\,\rm{L_{\odot}}$. As the non-detection in the optical bands and the presence of a massive molecular gas reservoir suggested, the SED modelling showed that the companion galaxy of SMM J04135+10277 is a highly dust-obscured starburst galaxy, reminiscent of submillimetre galaxies (SMGs).
\begin{table}
\caption{Results of the SED modelling\tablefootmark{a}}             
\label{table2}      
\centering          
\begin{tabular}{l c}
\hline\hline
\multicolumn{2}{c}{\rule{0pt}{2.5ex}\textsc{magphys} model\tablefootmark{b}}\\
\hline   
\rule{0pt}{2.5ex}log($M_{*}/\textrm{M}_{\odot}$) & $11.87^{+0.18}_{-0.22}$\\
\rule{0pt}{2.5ex}$\rm{SFR}\ [\textrm{M}_{\odot}/\textrm{yr}$] & $700^{+405}_{-315}$\\
\rule{0pt}{2.5ex}log($M_{\textrm{dust}}/\textrm{M}_{\odot}$) & $9.71^{+0.15}_{-0.17}$\\
\rule{0pt}{2.5ex}log($L_{\textrm{dust}}/\textrm{L}_{\odot}$) & $12.97^{+0.16}_{-0.15}$\\
\rule{0pt}{2.5ex}$T_{\textrm{dust}}/\textrm{K}$ & $34^{+9}_{-5}$\\
\rule{0pt}{2.5ex}$A_{V}$ [mag] & $2.8^{+0.4}_{-0.5}$\\     
\hline    
\end{tabular}
\tablefoot{
\tablefoottext{a}{The fitted parameters are corrected for gravitational lensing magnification ($\mu_{L}=1.3$; \citealt{2003A&A...411..343K}).}
\tablefoottext{b}{The lower and upper limits of the median values are given by the $16^{\textrm{th}}$ and $84^{\textrm{th}}$ percentile of the likelihood distribution.}
}
\end{table}
Compared to the ALESS SMGs \citep{2015ApJ...806..110D}, the estimated dust mass of $5.1\times 10^9\,\rm{M_{\odot}}$ is an order of magnitude higher, while the dust luminosity is higher by a factor of 2.
The estimated dust temperature of 34 K is consistent within uncertainties with the median dust temperature of the highest $A_{V}$ galaxies in the ALMA survey ($T_{\rm dust}=42\pm2\,{\rm K};\ A_{V}>1.9$ mag).
Although the  stellar mass is very high, it is not completely unexpected because of the very large molecular gas reservoir ($M_{\rm H_2} \sim 10^{11}$\,M$_\odot$, \citealt{2004ApJ...609...61H, 2011ApJ...739L..32R}).

However, since the resolution of SCUBA does not allow us to pinpoint the exact origin of the submm emission, the host galaxy of the quasar could have some contribution to the 850 $\mu$m emission. This means that the submm flux of the companion galaxy could be lower than what we assumed in our SED modelling, which in turn can influence the estimated properties of the companion galaxy. Future ALMA observations of SMM J04135+10277 will be able to resolve this system and put better constrain on its properties.
\section{How common is the case of SMM J04135+10277 at high redshift?}

\begin{figure*}
\begin{center}
\includegraphics[width=0.7\textwidth]{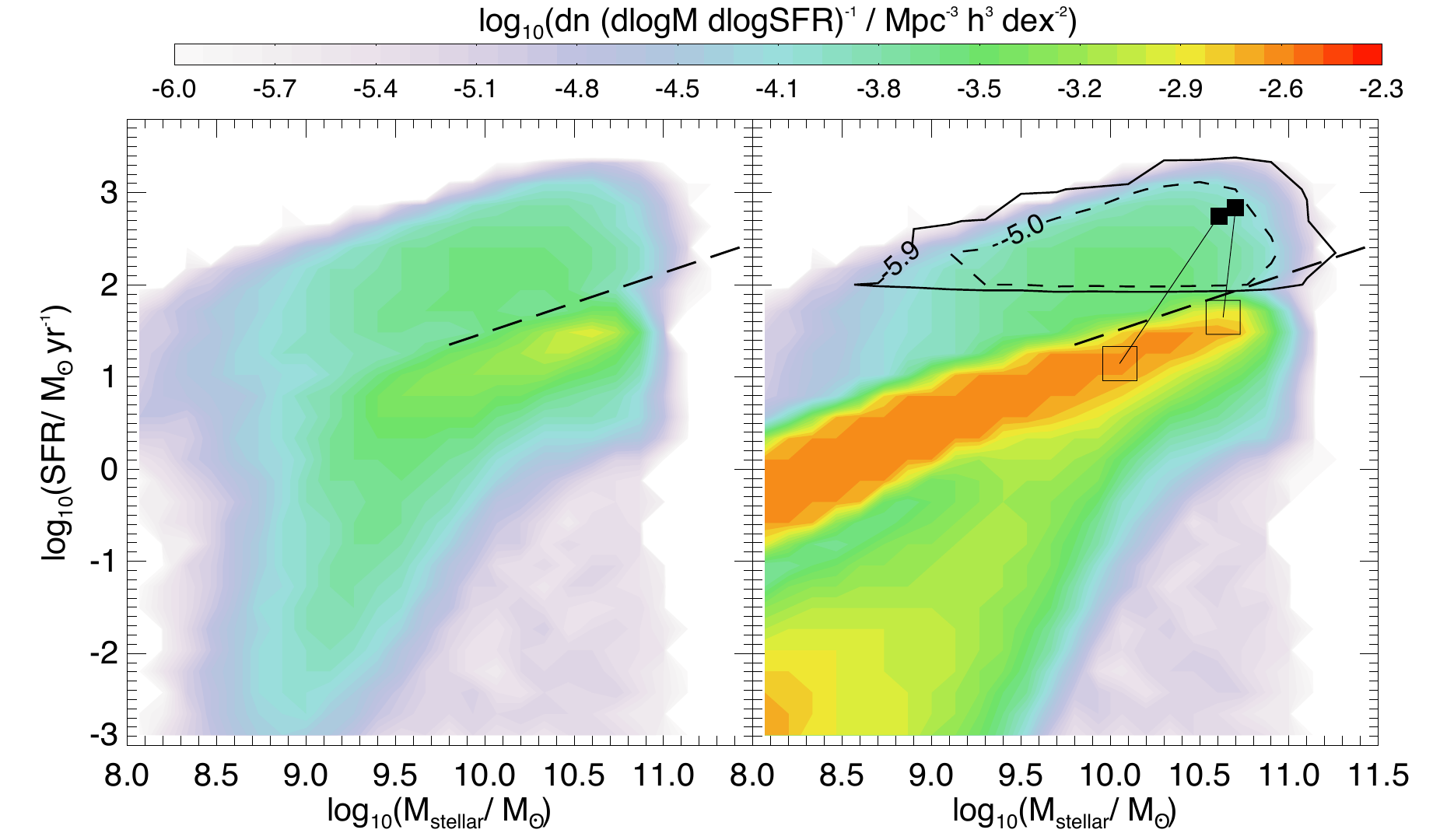}
\includegraphics[width=0.7\textwidth]{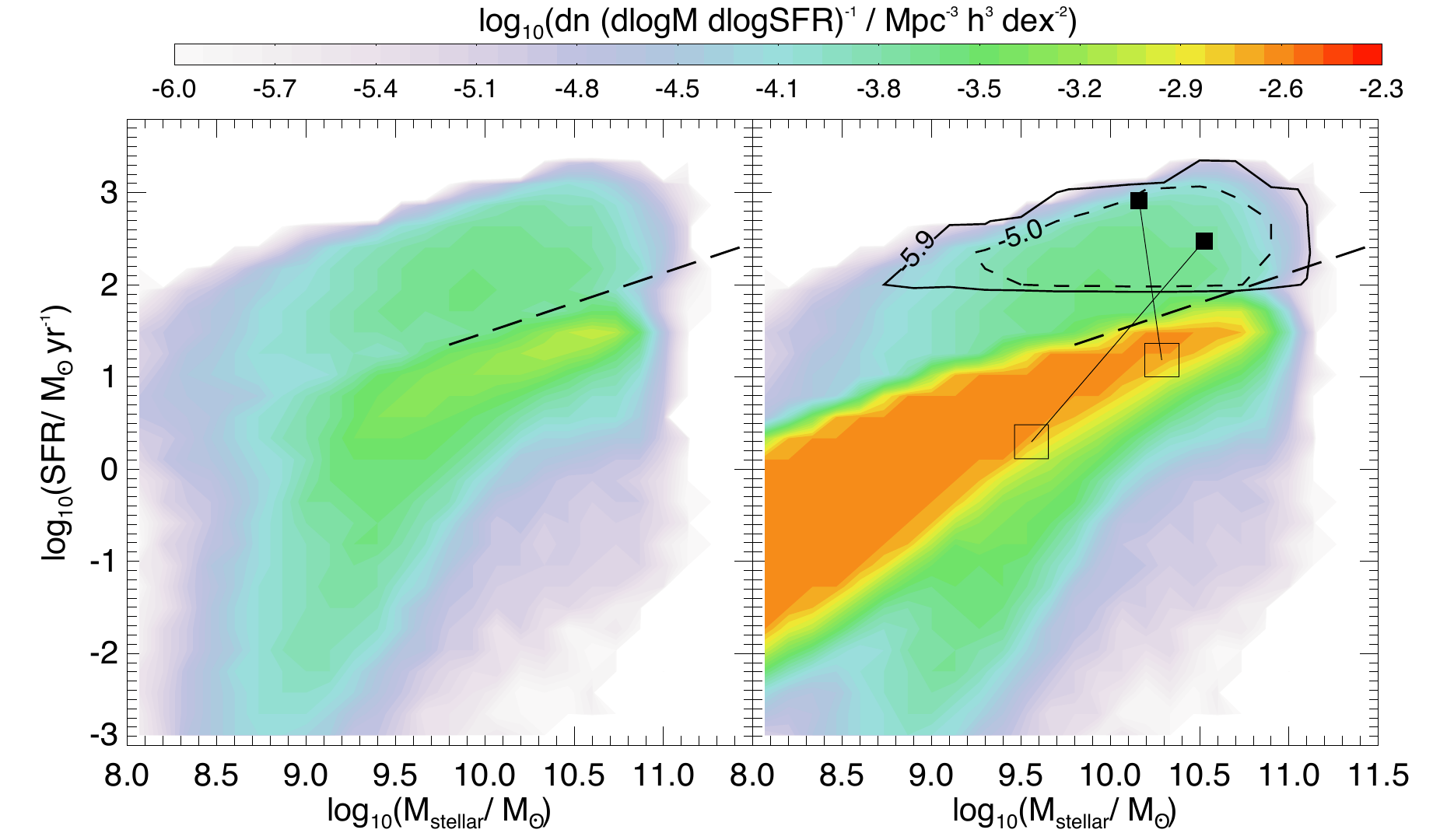}
\caption{{\it Top panels:} SFR-$M_{\star}$ plane at $z=2.8$ for the Gonzalez-Perez14 model.
 Both panels show as coloured contours the number densities of
 quasars with luminosities in the \textit{Bj} band $>10^{44.5}\,\rm erg\,s^{-1}$ (left panel) and of all galaxies in the model (right panel). Number densities correspond to the number per volume, per
 unit log$M_{\star}$ and logSFR (as indicated by the colour bar). The dashed black line indicates the main sequence of star-forming galaxies at a redshift $2<z<2.5$ \citep{2013ApJ...768...74T}. In addition, we show in the right panel as black contours the number density of all galaxies with $\rm SFR>100\,\rm M_{\odot}\, yr^{-1}$ that are at $<250$ kpc physical distance of a bright
 quasar ($L_{\rm B_j}>10^{44.5}\,\rm erg\,s^{-1}$). Solid and dashed contours show the number densities of $10^{-5.9}$ and $10^{-5}$$\,\rm Mpc^{-3}\,h^{3}\,dex^{-2}$. 
 These number densities are equivalent to $\approx 2$ and $\approx 16$ galaxies in the simulated volume, respectively. 
 This figure shows that the subset of highly star-forming galaxies living nearby
 bright quasars are about a tenth of all the highly star-forming galaxies. In the right panel we show two examples of quasar--companion galaxy pairs indicated by empty (quasar) and filled (companion) squares. These pairs are very similar to SMM J04135+10277, they have an integrated $\rm SFR>300\,\rm M_{\odot}\, yr^{-1}$ and the distance between the components is $<50$ kpc. {\it Bottom panels:} As in the top panels but for a variant of the Gonzalez-Perez14.GRP model.}
\label{fig4}
\end{center}
\end{figure*}
\begin{figure*}
\begin{center}
\includegraphics[width=0.7\textwidth]{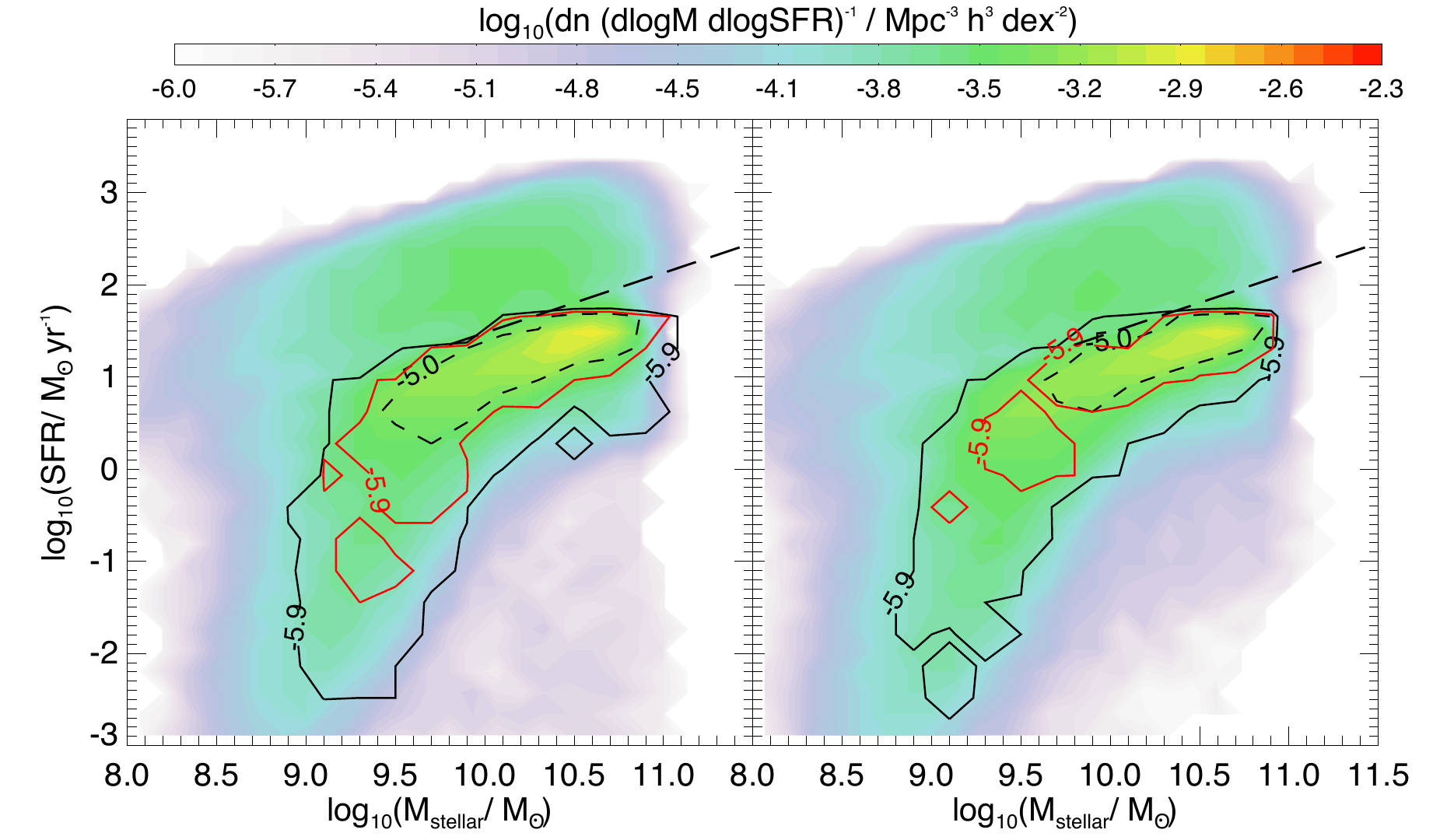}
\caption{{\it Left panel:} SFR-$M_{\star}$ plane at $z=2.8$ for the Gonzalez-Perez et al. (2014) model. The coloured contours show the number densities of quasars with luminosities in the \textit{Bj} band $>10^{44.5}\,\rm erg\,s^{-1}$. The black contours show the number densities of the subset of bright quasars that have at least one companion galaxy with $\rm SFR>100\,\rm M_{\odot}\,yr^{-1}$ at a physical distance $<1$~Mpc. The red contour shows the subset of bright quasars that have at least one companion galaxy with $\rm SFR>100\,\rm M_{\odot}\,yr^{-1}$ at a physical distance $<100$~kpc. Solid and dashed contours, and the dashed black line same as in Fig.~\ref{fig4}. 
The figure shows that the subset of bright quasars with nearby star-forming galaxies are typically on the main sequence of galaxies or below. {\it Right panel:} As in the left panel but for a variant of the  
 Gonzalez-Perez14.GRP model.}
\label{fig5}
\end{center}
\end{figure*}

With the aim of investigating the expected frequency of quasar--star-forming galaxy pairs in a cosmological set-up, we use {\sc galform}, which is a cosmological semi-analytic galaxy formation model \citep{2000MNRAS.319..168C}.
Since there are only a small number of known quasar--star-forming galaxy pairs, these simulations are essential to determine the expected frequency of such systems and their role in the evolution of galaxies. Moreover, the system of SMM J04135+10277 has been proposed to represent an early stage of a wet-dry merger event in which the host galaxy of the quasar is gas-poor or has significantly lower gas mass than the companion galaxy \citep{2013ApJ...765L..31R}; this is a unique feature among the known high-redshift AGN--companion galaxy systems. In the absence of high-resolution FIR surveys of high-redshift quasars, the only way to test the occurrence frequency of this system is to use numerical simulations.

\subsection{Simulation description}\label{simDesc}

\textsc{galform} uses analytic, physically motivated equations to follow the evolution of baryonic component and generate merger trees to trace the hierarchical structure formation governed by gravitational instability \citep{2006RPPh...69.3101B, 2010PhR...495...33B}. 
For this study we use the model described in \citet{2014MNRAS.439..264G}, which was developed using merger trees from the MS-W7 simulation \citep{Jiang14, 2016MNRAS.462.3854L}, which has a volume of $(500\, \rm Mpc\,h^{-1})^3$ and assumes the WMAP7 cosmology \citep{2011ApJS..192...18K}. The detailed description of the model can be found in \citet{2014MNRAS.439..264G} and \citet{2014MNRAS.440..920L}. Below we provide a brief explanation of the physical processes that are key to our investigation here.
\paragraph{Star formation.} 
\textsc{galform} adopts a star formation law for the disks of galaxies that relates the molecular gas surface density with 
the SFR surface density using a constant molecular gas depletion time, while during starbursts gas is converted into stars on a timescale that is proportional to the dynamical timescale of the bulge 
(model is described in detail in \citealt{Lagos11a} and \citealt{Lagos11b}). Thus, \textsc{galform} explicitely follows the partition of the gas into 
its atomic and molecular phases in both galaxy components, disk and bulge.
\paragraph{Black hole growth and AGN feedback.} The \textsc{galform}  model includes a sophisticated model of AGN feedback, 
black hole growth, and spin development; this model is described in detail in \citet{Fanidakis11} and \citet{Fanidakis12}. In short black holes (BHs) release energy through accretion of gas, making them visible as AGNs, and producing feedback. In \textsc{galform}, SMBHs grow in three ways: (i) accretion of gas during starbursts triggered by galaxy mergers or disk instabilities 
(starburst mode); (ii) accretion of gas from the hot halo (hot halo mode); and (iii) BH-BH mergers. The mass accreted onto the BH in a starburst is assumed to be a constant fraction of the mass
formed into stars, where the corresponding fraction is an adjustable parameter. The same processes that grow the mass of BHs can also change its spin. The radiative efficiency and the capability of 
a BH to launch relativistic jets are a function of the BH spin, mass, and accretion rate. Based on the latter model we calculate the optical and radio luminosity of all AGN in \textsc{galform}.
\paragraph{Satellite orbits and dynamical friction.} When galaxies become satellites (i.e. cross the virial radius of a more massive galaxy halo), 
their orbits are tagged to that of the sub-halos that host them, until the sub-halo gets stripped. This usually happens when the sub-halo is close to the centre of the host halo. Once the sub-halos cannot be tracked anymore, 
we assume the orbits of the satellite galaxies gradually decay towards the centre owing to energy and angular momentum losses driven by dynamical friction with the halo material \citep{Lacey93}.
\paragraph{Processing of hot gas in satellite galaxies.} The \textsc{galform} model assumes that when galaxies become satellites (i.e. cross the virial radius of a bigger halo) they instantaneously lose their 
hot gas reservoir to the hot gas of the central galaxy \citep{1980ApJ...237..692L, 2000ApJ...540..113B}. 
Here we also investigate a variant of this model that adopts gradual ram pressure stripping of the hot gas in satellite galaxies (i.e. satellite galaxies 
are able to retain their hot gas reservoir for longer; \citealt{2008MNRAS.383..593M, 2008MNRAS.389.1619F}). 
We refer to the \textsc{galform} model assuming instantaneous stripping as Gonzalez-Perez14 and the variant with gradual ram pressure stripping as Gonzalez-Perez14.GRP. \citet{Lagos14b} provide more details of this model.
\subsection{Quasar--star-forming galaxy pairs in \textsc{galform}}\label{simgalf}
We took the outputs of the simulation at $z=2.8$ to find quasar--companion galaxy pairs at different separations. We chose this redshift to compare the predictions of the simulation 
with the case of SMM J04135+10277. Specifically, we would like to determine the probability of finding such systems 
at this redshift and thus place SMM J04135+10277 in the context of the general galaxy and quasar population at 
$z=2.8$. The time slice in the simulations is $\sim200\,\rm Myr$, corresponding to $\Delta z\sim0.2$. We performed the calculations using both the Gonzalez-Perez14 and Gonzalez-Perez14.GRP models.
\paragraph{Characterising the quasar-galaxy pairs in {\sc galform}.}
 We created a simulated quasar sample at $z=2.8$ by taking all the simulated galaxies that have an AGN with a luminosity $L_{Bj}>10^{44.5}~\rm erg\,s^{-1}$ (\citealt{Fanidakis12}; we refer to these quasars as bright quasars) similar to SMM J04135+10277 ($L_{Bj}\sim2\times10^{45}~\rm erg\,s^{-1}$). In the simulated volume the number of bright quasars is $175,737$ and $182,147$ in case of the Gonzalez-Perez14 and the Gonzalez-Perez14.GRP models, respectively.
We selected all galaxies with $M_{\star}>10^8\,\rm M_{\odot}$ that reside at a physical distance $<1\,\rm Mpc$ from the quasar and named them companion galaxies (we also calculate the frequency of quasars with more massive companions; this is  
$M_{\star}>10^9\,\rm M_{\odot}$ and $M_{\star}>10^{10}\,\rm M_{\odot}$). We pay special attention to those companion 
galaxies that have $\rm{SFR}>100\,\rm{M_{\odot}\,\rm yr^{-1}}$, which we refer to as bright companions. 
Fig.~\ref{fig4} shows the SFR-$M_{\star}$ plane for our quasar sample and for all galaxies in the Gonzalez-Perez14 model at $z=2.8$. We also show the number densities of highly star-forming galaxies ($\rm{SFR}>100\,\rm{M_{\odot}\,\rm yr^{-1}}$) 
that reside at a physical distance $<250$~kpc of a quasar (right panel). 
Highly star-forming galaxies with a nearby quasar exclusively are about a tenth of the entire highly star-forming galaxy population, 
indicating that this phenomenon is not rare in the semi-analytic model.
Interestingly, we find that 80\% of our quasar sample that have bright companions at $<1$~Mpc are on or below the main sequence\footnote{It is a well-known problem of semi-analytical models and hydro-dynamical simulations that the main sequence is a factor of 2 offset from the observed main sequence, while they are able to reproduce the evolution of the density of stellar mass. For further details, see, e.g. \citet{2014MNRAS.444.2637M, 2015ApJ...799..201W}.}, which  
leads to the companion galaxy having a higher star formation rate and gas content than the quasar host itself in almost all of the 
cases in the simulation. We come 
back to this point later.

\begin{figure}
\begin{center}
\includegraphics[width=0.48\textwidth]{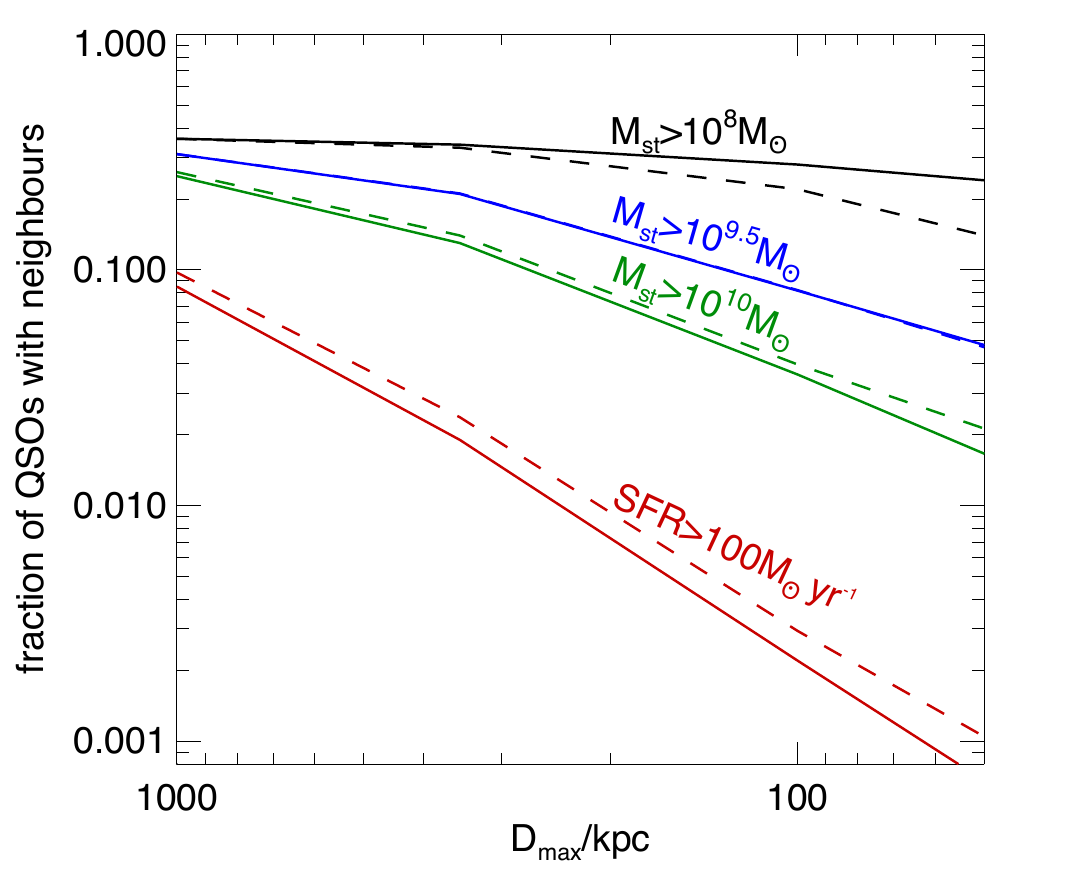}
\caption{Fraction of quasars with companions as a function of the aperture used for the search. The black, blue and green lines show this fraction in case of companions with different stellar masses ($M_{\star}>10^{8}\,\rm{M_{\odot}}$, $>10^{9.5}\,\rm{M_{\odot}}$, $>10^{10}\,\rm{M_{\odot}}$, respectively), while the red lines represent the fraction of quasars with bright companions ($\rm{SFR}>100\ \rm{M_{\odot}\,yr^{-1}}$) as a function of aperture, in the case of the
Gonzalez-Perez14 (solid lines) and Gonzalez-Perez14.GRP (dashed lines) models.}
\label{fig6}
\end{center}
\end{figure}

Both the Gonzalez-Perez14 and Gonzalez-Perez14.GRP models result in a very similar quasar simulated sample at $z=2.8$, but this is 
not true when taking all the galaxies in the simulation into account. In the right panels of 
Fig.~\ref{fig4} one can clearly appreciate the effect the model of hot gas stripping of satellite galaxies has on the galaxies that are below the main sequence of galaxies. 
In fact, we see that many of the galaxies that are passive (i.e. lie below the main sequence) in the Gonzalez-Perez14 model become normal star-forming galaxies (i.e. lie on the main sequence) in the Gonzalez-Perez14.GRP model (see \citealt{2014MNRAS.440..920L} and \citealt{2016MNRAS.461.3457G} for analyses of the effect on the 
gas content and the frequency of passive galaxies).
\paragraph{Frequency of quasar--companion galaxy systems at $z=2.8$.}
At a physical distance of $<1$~Mpc, $36$\% of quasars have at least one companion galaxy, while $10$\% of quasars have bright 
companions at $<1$~Mpc, showing how common this phenomenon is (Table~\ref{table3}). 
For companion galaxies of $M_{\star}>10^9\,\rm M_{\odot}$ and  $M_{\star}>10^{10}\,\rm M_{\odot}$, we find that $31$\% and $26$\% of 
quasars have at least one companion galaxy at a physical separation $<1$~Mpc, respectively.
Moving towards smaller separations between the quasar and its companions, the rates above decrease (see Fig.~\ref{fig6}).
At a physical distance $<100$~kpc, $22$\% of the quasars have companions and $0.3$\% have bright companions.
\paragraph{Gas content and star formation rate.}
Motivated by the finding of the CARMA CO(3-2) observations, which showed  
that for SMM J04135+10277 all the previously detected extended molecular gas reservoir is associated with a companion galaxy rather than the quasar host, 
we also studied the subset of quasars in the simulation whose host has a smaller molecular ($M_{\rm{H_{2}}}$) and atomic gas ($M_{\rm{HI}}$) 
reservoir than the integrated value of $M_{\rm{H_{2}}}$ and $M_{\rm{HI}}$ of all the companions. 
This rate was calculated using different apertures around the quasars. 
Following the results from CARMA, we speculate that the SCUBA detected submm emission (indicative of star formation) also comes from the companion galaxy of SMM J04135+10277 and not the quasar host.
With this in mind, we also calculated the number of quasars whose host has a lower SFR than the integrated value of all the companions at different apertures.

First we selected all quasar--companion galaxy pairs and we found that in $35$\% of the cases the $M_{\rm{H_{2}}}$ of the quasar host is smaller than the integrated value of all the companions 
at $<100$~kpc  distance. 
Moreover, in 37\% of the pairs the $M_{\rm{HI}}$ of the quasar host galaxy is also smaller than the integrated value of all the companions.
By tracking the distribution of the SFR in apertures around quasars, we find that in $40$\% of the pairs, 
the SFR of the quasar host is lower compared to the integrated value of its companions at a separation 
of $<100$~kpc.
Although we considered the integrated $M_{\rm{H_{2}}}$, $M_{\rm{HI}}$ and SFR of all companions, 
at small separations (e.g. $100$~kpc) the majority of quasars in the model 
have only one companion and, thus, this would observationally appear 
very similar to the SMM J04135+10277 system.
\subsection{The case of SMM J04135+10277 in the simulations}\label{simsmm}
The distance between the quasar and its companion galaxy in the system of SMM J04135+10277 is $\sim40$~kpc \citep{2013ApJ...765L..31R}. In the simulation the smallest median separation between a quasar and a companion galaxy is $\sim 30 \rm\, kpc$; below this distance it becomes difficult to properly resolve the distances. This distance is determined by the transition of satellite galaxies from having resolved halos to losing their halos from stripping. In our simulation the fraction of quasars that have companion galaxies at a distance $<50$~ kpc is $14$\% and $0.1$\% of our quasar sample have bright, star-forming companions at this distance. 

In order to make a one-to-one comparison between the model predictions and SMM J04135+10277, we focus on the subset of quasar--star-forming companion systems that 
have an integrated SFR inside the aperture $>500 \,\rm{M_{\odot}\,yr^{-1}}$ and compute their gas content and SFR.
We find that the percentage of quasars whose host has a smaller $M_{\rm{H_{2}}}$ and $M_{\rm{HI}}$ compared to the integrated value of their companions is $72$\% and $40$\% at a distance of $<100$~kpc and $67$\% and $35$\% at a distance of $<50$~kpc, respectively. This means, that according to our simulation a substantial part of the molecular gas of these systems lies in the companion galaxy and not in the quasar host galaxy. This is similar to the characteristics of SMM J04135+10277.

Interestingly, the percentage of quasars whose host has a lower SFR than the integrated value of their companions increases to 100\% compared to all quasar--companion galaxy pairs.
This suggests that the case of SMM J04135+10277 not only happens in the simulation, but it is also representative 
of highly star-forming quasar--galaxy systems at $z=2.8$.
We summarise the fraction of quasar--galaxy companion systems and the calculated frequencies in Table~\ref{table3}.

\begin{table*}
\caption{Simulation results}
\label{table3}
\centering
\begin{tabular}{lcccc}
\hline
\hline
\multicolumn{5}{c}{\rule{0pt}{2.5ex}Instantaneous (Gradual) Ram Pressure Stripping}\\
\hline
\rule{0pt}{2.5ex}Distance & <1 Mpc & <350 kpc & <100 kpc & <50 kpc\\
Companion\tablefootmark{a} & 36\% (36\%) & 33\% (34\%) &22\% (28\%) & 14\% (24\%)\\ 
Companion\tablefootmark{b} & 31\% (31\%) & 21\% (21\%) &8.2\% (8.2\%) & 4.7\% (4.8\%)\\ 
Companion\tablefootmark{c} & 26\% (25\%) & 14\% (13\%) &4\% (3.6\%) & 2.2\% (1.7\%)\\ 
Bright companion\tablefootmark{d} & 10\% (9\%) & 2.4\% (2\%) & 0.3\% (0.25\%) & 0.1\% (0.1\%)\\ 

$M_{\rm{H_{2}}}$\tablefootmark{e} & 93\% (93\%) & 69\% (71\%) & 35\% (39\%) & 24\% (30\%)\\ 

$M_{\rm{HI}}$\tablefootmark{e} & 94\% (95\%) & 73\% (75\%) & 37\% (41\%) & 26\% (32\%)\\ 

$\rm SFR$\tablefootmark{e} & 96\% (96\%) & 78\% (79\%) & 40\% (43\%) & 28\% (33\%)\\ 
\hline
\multicolumn{5}{c}{\rule{0pt}{2.5ex}quasar--companion galaxy systems with $\rm{SFR}_{\rm{integrated}}>500 \,\rm{M_{\odot}\,yr^{-1}}$}\\
\hline
\rule{0pt}{2.5ex}$M_{\rm{H_{2}}}$\tablefootmark{f} & 100\% (100\%) & 95\% (98\%) & 72\% (72\%) & 67\% (67\%)\\ 

$M_{\rm{HI}}$\tablefootmark{f} & 100\% (100\%) & 88\% (91\%) & 40\% (67\%) & 35\% (67\%)\\ 

$\rm SFR$\tablefootmark{f} & 100\% (100\%) & 100\% (100\%) & 100\% (100\%) & 100\% (100\%)\\
\hline
\hline
\end{tabular}
\tablefoot{
\tablefoottext{a}{Percentage of quasars ($L_{Bj}>10^{44.5}$ erg/s) that have companion galaxies with $M_{\star}>10^{8}\, \rm{M_{\odot}}$.}
\tablefoottext{b}{Percentage of quasars ($L_{Bj}>10^{44.5}$ erg/s) that have companion galaxies with $M_{\star}>10^{9}\, \rm{M_{\odot}}$.}
\tablefoottext{c}{Percentage of quasars ($L_{Bj}>10^{44.5}$ erg/s) that have companion galaxies with $M_{\star}>10^{10}\, \rm{M_{\odot}}$.}
\tablefoottext{d}{Percentage of quasars ($L_{Bj}>10^{44.5}$ erg/s) that have bright companion galaxies ($\rm{SFR}>100\, \rm{M_{\odot}\,yr^{-1}}$).}
\tablefoottext{e}{Percentage of quasar--companion galaxy pairs where the $\rm{H_{2}}$ mass, HI mass and the SFR of the quasar host is smaller than that of the companion galaxies. Here the rates are calculated from the sample of quasar--companion galaxy pairs at the given spherical aperture.}
\tablefoottext{f}{The same as in \tablefoottext{e}{} but in case of quasar--companion galaxy pairs that have an integrated $\rm{SFR}>500\,\rm{M_{\odot}\,yr^{-1}}$ inside the aperture.}
}
\end{table*}
\section{Discussion}

Compared to the few known quasar--star-forming companion galaxy systems, the case of SMM J04135+10277 is peculiar in the sense that the molecular gas mass of the companion galaxy sigficantly exceeds that of the host galaxy of the quasar. 
For example, the $z=4.7$ source BR 1202-0725 is fairly similar to the system of SMM J04135+10277 at first glance: it contains an optically luminous quasar (SE component), an obscured, submm bright companion galaxy (NW component), and probably two other $\rm Ly\alpha$-emitting galaxies \citep{2013ApJ...763..120C}. Both components have high FIR luminosities ($L_{\rm FIR}\sim10^{13}\,\rm L_{\odot}$; \citealt{2006ApJ...645L..97I}), indicating SFRs higher than $1000\,\rm M_{\odot}\,\rm yr^{-1}$ \citep{2006ApJ...645L..97I, 2012A&A...545A..57S, 2013A&A...559A..29C}. Although the SFR and FIR luminosity of the companion galaxy of SMM J04135+10277 is comparable to the NW component of BR 1202-0725, it has an order of magnitude higher dust mass and molecular gas mass. Moreover, the SE component of BR 1202-0725 is gas rich, while the host galaxy of the quasar in the system of SMM J04135+10277 seems to be gas poor.

A more similar case to SMM J04135+10277 was reported by \citet{2009ApJ...698L.188C}, who observed the quasar SDSS 160705+533558 ($z=3.65$) using the Submillimeter Array at $850\,\mu\rm  m$. \citet{2009ApJ...698L.188C} found extended submm emission ($10-35$~kpc) offset from the optical position of the quasar by $10$~kpc and concluded that this object might represent an early stage of a merger event between the gas-poor host galaxy of the AGN and a likely gas-rich, starburst galaxy with a $\rm SFR\sim3000-8000\,\rm M_{\odot}\,\rm yr^{-1}$ and infrared luminosity $L_{\rm IR}\sim2-5\times10^{13}\,\rm L_{\odot}$. Given the high SFR and infrared luminosity of this starburst galaxy, this case might be considered a scaled-up analogue of SMM J04135+10277.

The main motivation and aim of our simulations was to determine the uniqueness of systems similar to SMM J04135+10277.
Based on our simulations performed at $z=2.8$, 36\% of quasars have at least one companion galaxy in their $<1$ Mpc radius environment and 10\% of our simulated quasar sample have bright companion galaxies ($\rm{SFR}>100\ \rm{M_{\odot}\,yr^{-1}}$). This is in agreement with observational results found in the literature for high-$z$ AGNs and quasars.
For example, an order of magnitude overdensity of luminous, star-forming galaxies was found in the 3.5 arcmin$^2$ field around a $z=1.7$ quasar by SCUBA \citep{2004ApJ...604L..17S}, \citet{2008MNRAS.383..289P} discovered an excess of submm sources around $z>5$ quasars, and \citet{2010MNRAS.405.2623S} reported higher number counts of SMGs in the 2 arcmin diameter field of  five  X-ray selected quasars compared to blank fields ($1.7<z<2.8$).
Such overdensity was discovered in case of HzRGs as well \citep[e.g.][]{2000ApJ...542...27I, 2003ApJ...583..551S, 2013ApJ...769...79W}.

\citet{2015MNRAS.448.3325J} observed 30 obscured, presumably AGN-dominated galaxies with JMCT SCUBA-2 and 
found a factor of 4--6 overdensity of submm sources in the 1.5 arcmin radius reach of the primary galaxies compared to blank-field surveys. The sample was selected based on their extremely red mid-infrared SEDs and compact radio counterparts from the WISE all-sky infrared survey and NVSS/FIRST radio survey.
This suggests that at a scale of $\sim10\ \rm{Mpc}$, WISE/radio-selected AGNs are located in overdense environments and might be useful as tracers of high-density regions of dusty galaxies ($z=2-3$).

At smaller scales, at $<350$ kpc distance, our simulations show that 33\% of the quasar sample have a companion galaxy and 2.4\% have bright companions, while at a distance of less than 100 kpc these fractions are lower: 22\% of the quasar sample have a companion and only 0.3\% have bright companions at $z=2.8$. At even smaller scales ($<50$ kpc), such as in case of SMM J04135+10277, 14\% of our quasar sample have a close companion galaxy and 0.1\% have a bright, star-forming companion ($\rm{SFR}>100\ \rm{M_{\odot}/yr}$). This fraction is not neglegible and might be important in terms of galaxy evolution, especially when focusing on systems with high integrated SFR ($\gtrsim 500\,\rm{M_{\odot}\,yr^{-1}}$). In most of these quasar--star-forming pairs in the simulation, the star formation is dominated by the companion galaxy, a scenario similar to SMM J04135+10277. 
As some of these close systems ($<50$~kpc) will merge by $z=0$, they are probably the progenitors of local, massive ellipticals. The importance of our simulation is that it provides important clues about the formation of massive galaxies as we can find and study their progenitors.

The discovery of close quasar--star-forming galaxy systems is challenging, since it requires high-resolution observations. But as it has been seen in case of SMGs and dusty star-forming galaxies (DSFG), where a significant fraction of these sources resolved into two or more fainter components \citep{2012ApJ...761...89B, 2013ApJ...768...91H, 2013MNRAS.432....2K, 2015A&A...577A..29M, 2015ApJ...812...43B}, FIR-bright, \textit{Herschel} detected quasars could have unresolved companions as well.
As our simulation showed, in more than $70\%$ of quasar--companion-galaxy systems the host galaxy of the quasar has lower SFR than its companion galaxies at $<350$ kpc and, in case of systems with high integrated SFR inside the aperture ($\rm SFR>500 \,\rm{M_{\odot}\,yr^{-1}}$) this rate is $100\%$ (Table \ref{table3}). Therefore FIR observations of quasars without sufficiently high resolution could be misleading and the presence and contribution of companion galaxies to the FIR emission cannot be excluded.

Indeed, at a scale of 150 kpc \citet{2015ApJ...806L..25S} found a factor of 10 overdensity of  neighbour galaxies around WISE/NVSS selected dusty, luminous quasars at $z\sim2$, detected by ALMA Band 7, compared to unbiased regions. Thirty-five percent of their sample have companions within the ALMA primary beam and no trend was found between the presence of the neighbours and the redshift, total luminosity, 870 $\mu$m flux, and radio power of the central, dusty quasar.

Moreover, in the last decade several bright, gas-rich companions of quasars and HzRGs have been discovered but typically at higher redshifts ($z>2.8$) \citep[e.g.][]{2005A&A...430L...1D, 2008MNRAS.390.1117I, 2010MNRAS.404..198I, 2012MNRAS.425.1320I, 2013MNRAS.428.3206W, 2015MNRAS.452.2388H}. Thus the case of SMM J04135+10277 is interesting but not stand-alone and in the coming years more and more of these AGN--companion-galaxy systems will reveal themselves thanks to the attainable resolution and sensitivity in our era. 
\section{Conclusions}
In this paper we presented observational and modelling results of the gas-rich companion galaxy of the high-redshift quasar ($z=2.84$) SMM J04135+10277. This galaxy is particularly interesting, since it hosts one of the most massive molecular gas reservoirs found in the high-$z$ Universe ($M_{\rm H_2} \sim 10^{11}$\,M$_\odot$).
In order to investigate the expected frequency of such systems, we used the results of a galaxy formation and evolution model, \textsc{galform}. This model encapsulates the fundamental processes of galaxy evolution in simple, analytical equations, including the growth of black holes. We summarise our main results as follows.
\begin{itemize}
\item We constructed the SED of the companion galaxy using archive optical, infrared, and submm data and APEX ArTeMiS 350 $\mu$m observations. We modelled the SED using the \textsc{magphys} code with high-$z$ extensions \citep{2008MNRAS.388.1595D,2015ApJ...806..110D}. Based on our study the companion galaxy is a highly dust-obscured, starburst galaxy with a visual extinction of $A_{V}=2.8^{+0.4}_{-0.5}$ mag and SFR of $700^{+405}_{-315}\, \rm{M_{\odot}\,yr^{-1}}$.\\
\item {The estimated dust mass and dust luminosity of the companion galaxy is $5.1^{+2.1}_{-1.7}\times 10^9\, \rm{M_{\odot}}$ and $\textrm 9.3^{+4.2}_{-2.7} \times 10^{12}\, \rm{L_{\odot}}$, respectively. Both the dust mass and dust luminosity are higher compared to the average values found for the ALESS SMGs \citep{2015ApJ...806..110D}. The estimated dust temperature ($T_{\rm dust}=34^{+9}_{-5}$ K) is consistent within uncertainties with the median dust temperature of the highest $A_{V}$ galaxies in the ALMA survey.}\\
\item According to our simulation at a distance of $<350$ kpc 33\% of our quasar sample have a companion galaxy ($M_{\star}>10^8\, \rm{M_{\odot}}$) and 2.4\% have bright companions ($\rm{SFR}>100\,\rm{M_{\odot}\,yr^{-1}}$). At smaller scales ($<50$ kpc), such as in case of SMM J04135+10277, 14\% of our quasar sample have companion galaxies and 0.1\% have a bright, star-forming companion. These results show that the presence of galaxies around quasars is a common phenomenom and the presence of companion galaxies with high SFR is possible.\\
\item In two-third of quasar--star-forming galaxy companion systems that have an integrated SFR of $>500 \,\rm{M_{\odot}\,yr^{-1}}$ inside an aperture of $<50$~kpc, the companion galaxy has higher molecular gas mass and in all of these systems the companion galaxy dominates the SFR, just as in the case of SMM J04135+10277.\\
\item We compared our simulation results with observations of high-$z$ quasars and HzRGs and we found that our predictions are in agreement with the observations both at large and small scales. Our results suggest that quasar--gas-rich companion galaxy systems are common phenomena in the early Universe and the high incidence of companions makes the study of such systems crucial to understand the growth and hierarchical build-up of galaxies and black holes.
\end{itemize}
Identification of close pairs of quasars and star-forming galaxies, such as studied in this paper, is challenging. While sensitive single-dish telescopes have identified many FIR-bright quasars, future high angular resolution observations with, for example ALMA, will be necessary to quantify how large a percentage of the star formation takes place in the quasar host compared to companion galaxies.
\begin{acknowledgements}
We thank the anonymous referee for helpful suggestions and comments. We thank to Philippe Andr{\'e} for providing IDL data reduction scripts and Vera K{\"o}nyves for her help during the ArTeMis data reduction. J.F. would like to thank Lukas Lindroos for useful discussions.  J.F. and K.KK. acknowledge support from the Knut and Alice Wallenberg Foundation. K.K.K. acknowledges support from the Swedish Research Council.
CL is funded by a Discovery Early Career Researcher Award (DE150100618). VGP acknowledges support from a European Research
Council Starting Grant (DEGAS-259586). This study is based on observations with the Atacama Pathfinder EXperiment (APEX) telescope under programme ID 094.F-9336(A). APEX is a collaboration between the Max Planck Institute for Radio Astronomy, the European Southern Observatory, and the Onsala Space Observatory.
This work is based on observations made with the NASA/ESA Hubble Space Telescope, and obtained from the Hubble Legacy Archive, which is a collaboration between the Space Telescope Science Institute (STScI/NASA), the Space Telescope European Coordinating Facility (ST-ECF/ESA), and the Canadian Astronomy Data Centre (CADC/NRC/CSA). 
This work is based in part on observations made with the Spitzer Space Telescope, which is operated by the Jet Propulsion Laboratory, California Institute of Technology under a contract with NASA.
This research has made use of the NASA/ IPAC Infrared Science Archive, which is operated by the Jet Propulsion Laboratory, California Institute of Technology, under contract with the National Aeronautics and Space Administration.
This research used the facilities of the Canadian Astronomy Data Centre operated by the National Research Council of Canada with the support of the Canadian Space Agency.
Based on observations obtained with MegaPrime/MegaCam, a joint project of CFHT and CEA/DAPNIA, at the Canada-France-Hawaii Telescope (CFHT) which is operated by the National Research Council (NRC) of Canada, the Institut National des Science de l'Univers of the Centre National de la Recherche Scientifique (CNRS) of France, and the University of Hawaii.
The simulation work presented used the DiRAC Data Centric system at Durham University, operated by the Institute for Computational Cosmology on behalf of the STFC DiRAC HPC Facility ({\tt www.dirac.ac.uk}). This equipment was funded by BIS National E-infrastructure capital grant ST/K00042X/1, STFC capital grant ST/H008519/1, and STFC DiRAC Operations grant ST/K003267/1 and Durham University. DiRAC is part of the National E-Infrastructure.
\end{acknowledgements}
\bibliographystyle{aa}
\bibliography{fogasy_aa}

\end{document}